\documentstyle[12pt,worldsci]{article}
\newcommand{\Avg}[1]{\left\langle #1 \right\rangle}

\def\etal{{\it et.al.}\/}

\def\etc{{\it etc.}\/}

\def\eg{{\it e.g.}\/}
\def\vev{{\it v.e.v.}\/}

\def\ol#1{\overline{#1}}
\def\tw#1{\tilde{#1}}

\def\hf{\frac{1}{2}}
\def\roughlyup#1{\mathrel{\raise.3ex\hbox{$\sim$\kern-.75em
\lower1ex\hbox{$#1$}}}}
\def\roughlydown#1{\mathrel{\raise.3ex\hbox{$#1$\kern-.75em
\lower1ex\hbox{$\sim$}}}}
\def\bra{\langle}
\def\ket{\rangle}
\def\dket{\rangle\!\rangle}
\def\dbra{\langle\!\langle}
\def\lsim{\roughlydown<}
\def\gsim{\roughlydown>}
\def\simeq{\roughlyup-}
\def\hc{{h.c.}}

\def\scrs{\scriptscriptstyle}

\def\bfc#1{{{\bf #1}}}
\def\Sc#1{{{\cal #1}}}
\def\ss#1{{{\scriptscriptstyle #1}}}

\def\eqa{\begin{eqnarray}}
\def\eeqa{\end{eqnarray}}
\def\eq{\begin{equation}}
\def\eeq{\end{equation}}
\def\nn{\nonumber}

\def\Pl{\gamma_\lft}
\def\lft{{\scrs L}}
\def\rht{{\scrs R}}
\def\Pr{\gamma_\rht}
\def\veps{\varepsilon}
\def\eps{\epsilon}

\def\SM{{\scrs SM}}

\def\GF{G_{\scrs F}}

\def\GR{{\scrs GR}}
\def\bb{\beta\beta}
\def\bbtn{\bb_{2\nu}}
\def\bbm{\bb_{\varphi}}
\def\bbmm{\bb_{\varphi\varphi}}
\def\bbzn{\bb_{0\nu}}

\def\mi{m_{\nu_i}}
\def\mj{m_{\nu_j}}
\def\ma{m_{\ss{N}_a}}
\def\wf{w_{\scrs F}}
\def\wgt{w_{\scrs GT}}
\def\geff{g_{\rm eff}}
\def\bfrnm{{\bf r}_{\!nm}}
\def\gv{g_{\scrs V}}
\def\ga{g_{\scrs A}}
\def\geff{g_{\rm eff}}

\begin{document}

\rightline{McGill-95/13}
\rightline{hep-ph/9505349}

%\rightline{September 1994}

\title{{\bf SCALAR-EMITTING MODES IN DOUBLE-BETA DECAY}}
\author{C.P. BURGESS\thanks{Invited talk presented to the Workshop
on Double Beta Decay and Related Topics, Trento Italy, April 1995.} \\
\vspace{0.3cm}
%{\em Institut de Physique, Universit\'e de Neuch\^atel \\
%1 Rue A.L. Breguet, CH-2000 Neuch\^atel, Switzerland.}\\
%\vspace{0.2cm}
%and \\
%\vspace{0.2cm}
{\em
Physics Department, McGill University \\
3600 University St.,  Montr\'eal, Qu\'ebec, Canada, H3A 2T8.}\\
}
\maketitle
\setlength{\baselineskip}{2.6ex}

{\begin{center} ABSTRACT \end{center}
{\small \hspace*{0.3cm}
The sum-energy spectrum of electrons emitted in double-beta decay is a
well-known
diagnostic for the nature of the physics which is responsible for the decay.
Three types of spectra are usually considered when these experiments are
analysed:
one each for the standard two-neutrino ($\bbtn$) decay, neutrinoless ($\bbzn$)
and
a Majoron-emitting ($\bbm$) decay. It has recently been shown that two other
electron spectra can be possible for scalar-emitting modes, in addition to
these traditional three.  One of these is softer than the Standard-Model
$\bbtn$ decay, while the other is intermediate between the $\bbtn$ and the
usual $\bbm$ spectra. The
models which predict these new spectra are generically more natural than those
which predict the traditional $\bbm$ spectrum, in that they can
accomodate the constraints following from the steadily improving limits on
$\bbzn$
decay without requiring the fine-tuning that is endemic to the usual models.
This article reviews the properties of the physics which can produce the new
kinds
of electron spectra.   }}

\section{Introduction}

Double-beta ($\bb$) decay is an extremely rare process in which two nuclear
neutrons simultaneously convert into two protons and two electrons. Within the
Standard Electroweak Model (SM) this decay occurs at second order in the
charged-current weak interactions, and is accompanied by the emission of two
antineutrinos, giving rise to a characteristic ($\bbtn$) electron spectrum.
Despite the extremely long half-lives involved --- typically $10^{20}$ yr or
more --- heroic efforts\cite{expreview} over the past ten years have been
rewarded
by its experimental detection.

Because it is such a rare process, $\bb$ decay experiments also furnish a
unique
window onto whatever new physics may replace the SM at energies very much
higher
than those that are directly accessible in today's accelerators. This is
because
the effects for these experiments of new interactions can in some circumstances
compete with those of run-of-the-mill SM decays.

In order to be detectable in $\bb$ experiments, new physics must
have either or both of the following properties:

\begin{enumerate}
\item
It must violate a selection rule --- \eg: electron-number ($L_e$)
conservation --- which is satisfied by the SM contribution;

\item
It must contain new particle states that are light enough to be produced in
$\bb$
decay. Since $Q \sim 1$ MeV is typical of the energy release in these decays,
any
such new particles must be much lighter than this scale.
\end{enumerate}

The purpose of this article is to outline the features of types of new physics
which satisfy the second of these properties, producing $\bb$ decays in which
 new light particles are emitted. The title refers to
`scalar-emitting' decay modes because, with the occasional
exception\cite{vectors}, this new light particle is usually taken to be
spinless.
A spinless particle can be particularly well-motivated if it is a (possibly
pseudo-) Nambu-Goldstone boson (NGB), since in that case its small mass is
naturally
understood. In particular, the focus is on comparatively recent
work\cite{CMM,icnapp,multimajoron} which shows that models of this sort
generally
have very different features --- including qualitatively different experimental
signatures, such as electron spectra --- than are usually assumed in the
analyses
of current $\bb$ experiments.

\section{The Electron Energy Spectrum}

The experimental quantity that is used to distinguish exotic decays from the
ordinary SM events in $\bb$ experiments is the shape of the decay rate as a
function of the energies, $\veps_1$ and $\veps_2$, of the two emitted
electrons.
For instance, if no new light particles are produced, then $L_e$-violating new
physics can be identified if it produces decays which are `neutrinoless'
($\bbzn$)
in the sense that only electrons emerge from the decaying nucleus. In this case
the decay rate vanishes unless the sum of the two electron energies, $\veps =
\veps_1 + \veps_2$, equals the released energy, $Q$.

Decays into electrons plus light scalars, on the other hand, predict a
continuous
decay distribution throughout the entire kinematically allowed interval, $2 m_e
\le \veps \le Q$, that is distinguishable from both the SM $\bbtn$
distribution,
and the neutrinoless $\bbzn$ contribution at the endpoint, $\veps = Q$. Indeed,
the
prediction of the scalar-mediated ($\bbm$) decay in models like the
Gelmini-Roncadelli (GR) model of lepton-number breaking\cite{GR,GGN}
helped to motivate the original $\bb$ experiments.

\subsection{The Spectral Index}

The different spectra that are possible in the various decays can be
characterized
(except for $\bbzn$) in terms of a single integer\cite{CMM}, or `spectral
index', $n$.
This is because the decay distribution quite generally can be written in the
following form:
\eq
\label{spectrum}
{d \Gamma \over d\veps_1 d\veps_2} = \Gamma_0 \; (Q - \veps_1 -
\veps_2)^n \;  \left[ p_1 \veps_1 F(\veps_1) \right]  \; \left[ p_2
\veps_2 F(\veps_2) \right]  ,
\eeq
where $F(\veps_i)$ is the Fermi function which describes the distortion of the
decay distribution due to the electrostatic field of the nucleus, and $p_i
= |\bfc{p}_i|$, for $i=1,2$, represent the magnitudes of the three-momenta of
the
electrons. We neglect the mass, $\mu$, of the light scalar (or
scalars) that are emitted in writing eq.~(\ref{spectrum}). Should $\mu$ not be
negligible in comparison with $Q$, then the factor $(Q - \veps_1 - \veps_2)^n$
should be replaced by $[(Q - \veps_1 - \veps_2)^2 - \mu^2]^{n/2}$. Of course,
if there are several types of possible decay, the total spectrum is a sum of
terms like eq.~(\ref{spectrum}).

A plot of the spectral shape which follows from eq.~(\ref{spectrum}) for
various
choices for $n$, is given in Fig. 1.

The spectral shape is determined by $n$ because the normalization, $\Gamma_0$,
does not depend, to a very good approximation, on the two electron energies,
$\veps_1$ and $\veps_2$. This is  because the most important quantity which
sets the scale for contributions to $\Gamma_0$ is the typical momenta,
$p_{\ss{N}} \sim 60$ MeV, of  the decaying neutrons.  Since the sizes of
the light particle momenta are set by the net energy release, $Q \sim 1$ MeV,
they are negligible in their contribution to $\Gamma_0$, and in this
approximation
the spectral shape becomes determined purely by phase space and the Fermi
functions.

For example, the phase space of the two emitted neutrinos (plus the electrons)
of $\bbtn$ in the SM implies the corresponding spectral index is $n_\SM = 5$.
Using the phase space of  a single scalar instead of two neutrinos
similarly gives the result $n_\GR = 1$ for the spectrum predicted
for $\bbm$ decay by the GR model. With one early exception\cite{twoscalar},
this
same spectrum is also predicted by all of the alternative models for
scalar-emitting decays which have been proposed ever since the original GR
paper.

Of course, Nature need not be limited to the two choices $n=5$ and $n=1$, and
in general other values for $n$ might be expected to be possible signals for
$\bb$
experiments. It turns out that this naive expectation is true, and that models
exist which predict both\cite{twoscalar,CMM,multimajoron} $n=3$
and\cite{multimajoron} $n=7$.\footnote{As will become clear later, the
cases $n=5, 9,...$ can also be obtained by combining the features which
produce the $n=1, 3$ and $7$ decays.}
Furthermore, some of these models --- particularly
those for which $n=3$ --- can produce observable signals in $\bb$ experiments
while preserving agreement with all other constraints, such as those arising
from
precision electroweak measurements on the $Z$ resonance.

\vspace{0.08in}
\begin{center}
%%Begin InstantTeX Picture
\let\picnaturalsize=N
\def\picsize{2.3in}
\def\picfilename{multispectrum.ps}
%If you do not have the picture file add:
%\let\nopictures=Y
%to the beginning of the file.
\ifx\nopictures Y\else{\ifx\epsfloaded Y\else\input epsf \fi
\let\epsfloaded=Y
\centerline{\ifx\picnaturalsize N\epsfxsize \picsize\fi
\epsfbox{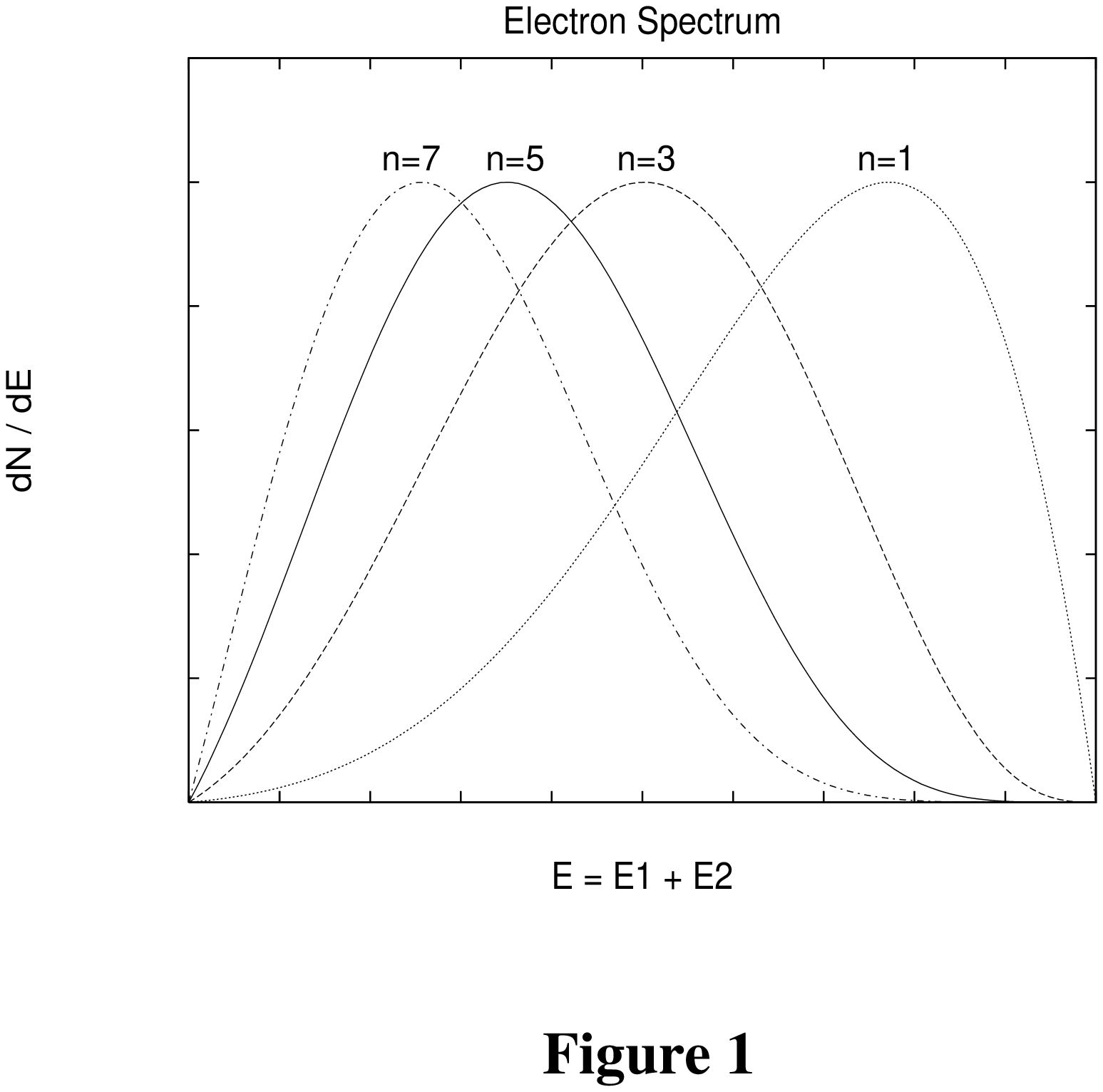}}}\fi
%%End InstantTeX Picture

%\medskip
%{\bf Figure 1}\\
%\medskip
$\hbox{}$
\vspace{1.7cm}
\end{center}

\begin{quote}
{\footnotesize The $\bb$ spectrum as a function of the two electrons' total
kinetic energy for various choices of the `spectral index' $n$. $n=1$
corresponds to the dotted line, $n=3$ is the dashed line, $n=5$ is the
solid line and $n=7$ is the dash-dotted line. All four curves have been
arbitrarily assigned the same maximal value for purposes of comparison.}
\end{quote}

\subsection{How to Generate $n \ne 1$}

The remainder of this article is intended to briefly outline the properties of
models with each of the values $n=1,3$ and $7$. Before doing so in detail,
however,
it is worth identifying the two general features which go into the prediction
of
the index $n$ for any model\cite{icnapp}. These are:

\begin{enumerate}
\item
{\sl Phase Space:} As was noted above, the phase space that is associated with
a
decay into two electrons and a scalar in $\bbm$ decay implies an index $n=1$,
as
for the GR model. The phase space of each additional scalar that appears in the
final state similarly increases $n$ by 2, so that in the absence of other
contributions to $n$, a two-scalar decay ($\bbmm$) should have: $n=3$, a
three-scalar decay: $n=5$, and so on.

\item
{\sl Nambu-Goldstone Bosons:} As is well known, if a scalar is a NGB, then
its couplings all must vanish in the limit of zero energy and
momentum. This generally implies a suppression of the contribution of
such a scalar in any low-energy process, and in particular for $\bb$ decay.
Since
this suppression implies that the emission amplitude for each NGB is
proportional
to a factor of the NGB momentum, every such particle in the $\bb$ final state
should also
increase $n$ by 2, in addition to the requirements of phase space. \footnote{
See,
however, section 2.3 below for a qualification to this statement.}
\end{enumerate}

It is immediately clear how to generate scalar-emitting $\bb$ decays for which
$n \ne 1$. If two scalars whose emission amplitude is not derivatively
suppressed
are produced in $\bbmm$ decay, then the index for the decay should be $n=3$. If
$N$ such scalars are emitted then $n=2N-1$. Alternatively, if a single scalar
is
emitted in a $\bbm$ event, but this scalar has the derivatively-suppressed
couplings of a NGB, then we again expect $n=3$. Similarly, if two such
derivatively-suppressed scalars are emitted, then the index for the the
corresponding decay is $n=7$. These arguments are borne out by detailed
calculations.\cite{multimajoron}

\subsection{A Puzzle with the GR Model}

A puzzle remains as to how the original GR model itself, and its many
successors
over the years, fit into the above counting scheme. After all, the light scalar
which is emitted in $\bbm$ decay in the GR model {\em is} a NGB: It is the NGB
for the spontaneous breaking of lepton number. Yet even so, the spectral index
which is predicted for this decay is $n_\GR=1$, rather than $n=3$ as the
previous
counting would have predicted. The resolution of this puzzle is instructive
because it reveals an additional criterion for increasing the spectral
index of a model.

The resolution to this puzzle\cite{CMM} goes as follows. It
is useful to think in terms of variables for which the NGB of the GR model is
explicitly
derivatively coupled. (These variables may be obtained from the standard ones
by
performing a field-dependent lepton-number rotation.) In these new
variables the NGB couples directly to the electron, as well as to the various
neutrinos of the model. The graphs which then dominate $\bbm$ decay at low
energies
turn out to be those for which the NGB of the final state is emitted from the
external electron lines, rather than from the neutrino lines.

But the emission of a massless boson with a vector coupling by an external
electron line introduces a potential singularity into the amplitude at low
momenta. Indeed if the emitted particle had been a photon rather than a NGB,
such
graphs would really be infrared divergent. For NGB emission, however, the
infrared singularity of these graphs is compensated by the derivative coupling
of
the massless scalar, giving a nonzero and finite result in the zero-momentum
limit.

In practice, then, in order to obtain a real suppression of $\bbm$ or $\bbmm$
decay because of the NGB nature of the emitted scalar, it is also necessary to
forbid the emission of the scalar from the external electrons in the decay. As
is shown below, this is an automatic feature of many models once they are
required to not produce experimentally unacceptable rates for $\bbzn$ decay in
a
natural way.

\section{Models: Preliminaries}

The remainder of the article is devoted to outlining the properties of the
various
kinds of scalar-emitting decays which the $\bb$ experiments can see. The
present
section sketches those features which are required as preliminaries to model
building, and the properties of some explicit models are briefly discussed in
the
section immediately following. Some of the more general
conclusions that can be drawn from a comparison of these models are finally
summarized in section 5.

\subsection{The Naturalness Problem}

The purpose of this section is to argue that a serious fine-tuning problem
exists for
virtually {\em all} of the models that have been proposed to date which predict
$n=1$ for the single-scalar $\bbm$ decays, or which predict $n=3$ for the
two-scalar
$\bbmm$ decays.  This fine tuning arises from a naturalness issue which has
strong implications for any model which purports to predict a detectably large
scalar-emitting $\bb$ decay rate.

This naturalness issue hinges on the following two questions which any such
model
must address:

\begin{enumerate}
\item
{\sl Masses:} The first question is: How can an elementary scalar have a mass
that is smaller than $Q \sim 1$ MeV, and so be light enough to permit its
production in $\bb$ decay? Being some five orders of magnitude lighter than the
electroweak scale, such a small scalar mass introduces a fine-tuning,
or `heirarchy', problem unless there is a mechanism which can protect it from
virtual effects at the weak scale.

\item
$\bbzn$ {\sl Decay:} The second question concerns the size that is predicted
for
$\bbzn$ decay. A model which breaks $L_e$ generically also predicts a nonzero
$\bbzn$ decay rate, and this rate must not be larger than the current
experimental
limit. Since this rate is often related to the rate for the scalar-emitting
$\bb$
decay, which is by assumption detectable, it can be difficult to suppress the
one
reaction without suppressing both. This makes the resulting constraint quite
powerful.
\end{enumerate}

There are several ways in which the various existing models handle these
issues.
Two mechanisms have been proposed which can permit such a naturally small
scalar
mass. Although (somewhat surprisingly) supersymmetry can be used to do the
job\cite{oscar}, the resulting models are quite complicated and fairly
contrived.
The much simpler alternative is to simply follow the original
workers\cite{GR,CMP}, and to require that the light scalar be the NGB for an
exact
or approximate global symmetry.

The second question becomes important if the symmetry for which the scalar is
the
NGB is electron number itself, as is the case for virtually all proposed models
which predict $n=1$ as the index for the scalar-emitting $\bb$ decays. This
includes essentially all models that have been considered until very
recently, and is one of the main motivations for taking the models with
$n\ne 1$ as important alternatives.

The problem arises because the same vacuum expectation value (\vev), $v$, which
breaks $L_e$ typically also generates a Majorana mass for the electron neutrino
whose size is $m_{\nu_e\nu_e} \sim \geff \,  v$, where $\geff$ is the relevant
scalar-neutrino Yukawa coupling constant. Such a Majorana mass gives rise
to $\bbzn$ decays which would have been seen if they exceed the current
experimental limit, which is $|m_{\nu_e\nu_e}| \lsim  1$ eV. But this limit
cannot
be satisfied simply by making $\geff$ very small, since if the $\bbm$ decay
itself
is to be observably large, then $\geff$ cannot be made smaller
than\cite{expgeff,kai,heid}  $\sim 10^{-4}$.  From this lower limit for $\geff$
we
learn that the $L_e$-breaking  \vev\ must satisfy $v \lsim 10$ keV.

But this upper bound requires the appearance in the scalar potential of a scale
that is more than 5 orders of magnitude smaller than the electroweak scale. As
such, it poses precisely the same type of fine-tuning problem that would have
occured if it were the scalar mass itself that was to be fine tuned. In either
case we are led to a mass scale in the scalar potential that is at largest
several hundred keV or so.

It is tempting to argue that a heirarchy of the size of order $v/M_{\scrs W}
\sim 100 \, \hbox{keV}/100 \, \hbox{GeV} \sim 10^{-6}$ is not so small if the
$\varphi - \nu$ coupling is really $\geff \sim 10^{-4}$, since in this case
radiative
corrections would be $\delta v \sim (\geff/4 \pi) \sim 10^{-5}$. Although
seductive,
this argument turns out to be wrong. The coupling, $\geff$, which appears in
$\bbm$
decay is not simply a yukawa coupling, $g$, of the lagrangian, but is really an
effective coupling in which a yukawa coupling is multiplied by a small mixing
angle:
$\geff \sim g \sin^2\theta$. The mixing angle arises because the precision data
at
LEP precludes the direct coupling of any new light scalar to the electroweak
sector. Scalar-emitting $\bb$ decay must therefore be induced through the
mixing
of an electroweak eigenstate with another state which couples to the light
scalar.
This mixing can occur in either the neutrino sector, giving sterile-neutrino
mediated decays, or, for example, in the scalar sector. In either case, the
mixing
angles typically cannot be too big without running into conflict with other
experiments. In sterile-neutrino models, for example, weak interaction bounds
imply that $\sin\theta$ cannot be larger than 10\% or so.

Now comes the main point. Because of its suppression by powers of
$\sin\theta$, $\geff$ can only be as large as $10^{-4}$ if the underlying
Yukawa coupling is considerably bigger. For example $\sin\theta \lsim 10 \%$
implies $g \gsim 10^{-2}$. But it is $g$ and not $\geff$ which controls the
radiative corrections to the scalar potential, so $\delta v$ is of order
$(g / 4 \pi)$ rather than $(\geff/4 \pi)$ as was argued above.  As a result
the estimate using $\geff$ underestimates the correction to $v$ by several
orders
of magnitude. More realistic estimates\cite{CMM,oscar} show that these
corrections can
only be sufficiently small in restrictive corners of parameters space, for
which new
degrees of freedom (like sterile neutrinos) are quite light, and so are
strongly
constrained phenomenologically.

To be sure, the $\bbzn$ decay rate could be made naturally small if $L_e$ were
conserved, and so if $v$ were to vanish. But in this case the corresponding NGB
disappears and the problem with the scalar mass must be solved in some other
way.

These considerations point to a natural way out of this dilemma. Both the
scalar mass and the $\bbzn$ decay rate can be naturally zero if: (a) the scalar
is a NGB, and (b) $L_e$ is {\em unbroken}. It follows that the light scalar
must
then be a NGB for some symmetry other than that responsible for electron-number
conservation. These two conditions are sufficient in themselves to imply a
spectral index for the associated scalar-emitting $\bb$ processes which is
respectively $n=3$ for $\bbm$, and $n=7$ for $\bbmm$ decay. This is easily seen
since these decays can only proceed if the emitted scalar itself carries
nonzero electron number which, together with $L_e$ and electric-charge
conservation, precludes the possibility of scalar emission from the
external electron lines.

\subsection{$\Gamma_0$ and Nuclear Form Factors}

Before turning to representative models for each type of decay, it is useful to
pause to record expressions for the normalization of the various $\bb$ decay
rates in a way which facilitates the comparison of different models.
These expressions are required in order to determine the kinds of couplings
which would be necessary in order to obtain observable scalar-emitting $\bb$
decay
rates.

It is useful to use the $\bb$ decay rate in the form given in
eq.~(\ref{spectrum}), with the normalization given by
\eq
\label{genericrate}
\Gamma_0(\bb_i) = {(\GF\cos\theta_{\scrs C})^4 \over 8(2\pi)^5} \; \left|
\Sc{A}(\bb_i) \right|^2 ,
\eeq
where $\GF$ is the Fermi constant, $\theta_{\scrs C}$ the Cabibbo angle,
and $\Sc{A}$ an amplitude which depends on the decay process being computed
($\bb_i$, with $i = 2\nu$, $\varphi$ or $\varphi\varphi$), on the couplings of
the
model, and on some soon-to-be-identified nuclear matrix elements.

$\Sc{A}$ can be written, for $0^+ \to 0^+$ transitions, as a
convolution:
\eq
\label{convolution}
\Sc{A} = \int {d^4\ell \over (2\pi)^4} \; L^{\mu\nu}(\ell) W_{\mu\nu}(\ell),
\eeq
where $L^{\mu\nu}$ depends on the detailed properties of the leptons in the
model, and $W_{\mu\nu}(\ell)$ contains the nuclear matrix elements that are
relevant to the decay:
\eq
\label{matrixelement}
W_{\mu\nu}(\ell) \equiv (2 \pi)^3 \, \sqrt{ { E E'\over M M'}} \;
\int d^4x \;\bra N'|T^* \left[ J_\mu(x) J_\nu(0) \right] | N\ket \;
e^{i\ell x} .
\eeq
Here $J_\mu = \bar{u} \gamma_\mu (1+\gamma_5) d$ is the weak charged
current that causes transitions from neutrons to protons, and $|N \ket$ and
$|N' \ket$ represent the initial and final $0^+$ nuclei in the decay. $E$ and
$M$ are the energy and mass of the initial nucleus, $N$, while $E'$ and $M'$
are the corresponding properties for the final nucleus, $N'$. The symmetries of
the problem ensure that the most general possible form for $W_{\mu\nu}$ is
\cite{CMM}:
\eqa
\label{formfactors}
W_{\mu\nu}(\ell) &=& w_1 \; \eta_{\mu\nu} + w_2 \; u_\mu
        u_\nu +w_3 \; \ell_\mu \ell_\nu + w_4 \; (\ell_\mu u_\nu +
                                        \ell_\nu u_\mu) \nn\\
    &&+ w_5 \; (\ell_\mu u_\nu - \ell_\nu u_\mu) +
     iw_6 \; \epsilon_{\mu\nu\sigma\rho}
     u^\sigma \ell^\rho ,
\eeqa
where $u_\mu$ is the four-velocity of the initial and final nucleus, and
the six Lorentz-invariant form factors, $w_a = w_a(u\cdot \ell, \ell^2), a =
1,\dots,6$, are functions of the two independent invariants that can be
constructed
from $\ell_\mu$ and $u_\mu$.

These form factors can be related\cite{CMM} to the nuclear matrix elements as
they are quoted in the literature\cite{DoiTomoda,Haxton,Rosen,KK}. For example,
in many situations (such as $\bbtn$, $\bbzn$ and some kinds of $\bbm$ and
$\bbmm$ decays)
$L^{\mu\nu} \propto \eta^{\mu\nu}$ and so the decay rate only depends on the
combination ${W^\mu}_\mu$. This is simply the difference between the
Fermi and Gamow-Teller form factors, which are defined by $\wf \equiv W_{00}$
and $\wgt \equiv \sum_{k=1}^3 W_{kk}$.  These form
factors, when computed using the closure and nonrelativistic impulse
approximations in a model of the nucleus, become
\eqa
\label{connection}
\wf &=& {2i \eps \gv^2\over \ell_0^2 - \eps^2 + i\varepsilon}
\; \dbra N'|\sum_{nm}e^{-i\bfc{l}\cdot\bfrnm}\tau^+_n \tau^+_m |N \dket;\nn\\
         \wgt &= &{2i \eps \ga^2\over \ell_0^2 - \eps^2 + i\varepsilon} \;
\dbra N'|\sum_{nm} e^{-i\bfc{l} \cdot \bfrnm} \tau^+_n\tau^+_m
        \; \vec\sigma_n \! \cdot\! \vec\sigma_m |N \dket .
 \eeqa
Here $\eps \equiv \ol{E} - M$ is the average excitation energy of the
intermediate nuclear state, $\bfrnm$ is the separation in position between
the two decaying nucleons, $\bfc{l}$ is the spatial component of $\ell^\mu$
in the nuclear rest frame, and $\gv \simeq 1$ and $\ga \simeq 1.25$ are the
vector and axial couplings of the nucleon to the weak currents. Finally,
$\dbra N'| \Sc{O} |N \dket$ represents a reduced matrix element from which
the nuclear centre-of-mass motion has been extracted, and $\tau^+_n$
(or $\vec\sigma_n$) are the raising operators for nuclear isospin (or the
nuclear spin operators) acting on the $n$'th nucleon.

\section{Models: Specific Examples}

We now turn to the details of representative models for which $n=1,3$ and 7.
Although much of what follows applies more generally we choose, for simplicity
and for the purposes of comparison, models in which the corresponding $\bb$
decay
proceeds due to the mixing of the electron neutrino, $\nu_e$, with a collection
of
sterile neutrinos, $N_i$.  The reader who is not interested in the specific
properties
of these models, can skip to the next, concluding, section in which their
general
features are compared.

\subsection{The Case $n=1$}

The simplest models to build are those for which the spectral index takes its
traditional value, $n=1$. Theories of this sort, when constructed using sterile
neutrinos, are similar to the original singlet-majoron models\cite{CMP}. They,
as well as other variants with $n=1$ which do not rely on sterile neutrinos to
produce $\bbm$ decays\cite{BSV}, have recently received renewed attention.

Suppose the neutrino mass eigenstates are denoted generically by $\nu_i$, and
their overlap with the electron-neutrino flavour eigenstate is called $V_{ei}$.
Suppose also that $\varphi$ represents the light scalar of the model. Then,
taking
the general Yukawa coupling lagrangian between these particles to be
 \eq
\label{yukawa}
\Sc{L}_{\varphi\nu\nu} = - \hf\; \bar{\nu}_i (a_{ij} \Pl  + b_{ij}  \Pr)
   \; \nu_j \; \varphi^* + c.c. \, ,
\eeq
and evaluating the Feynman graph of Fig. 2, gives an amplitude,\cite{CMM}
$\Sc{A}$ of the form of eq.~(\ref{convolution}), with\footnote{Our conventions
here use  $\eta^{\mu\nu} = \hbox{diag}(-+++)$.}
\eq
\label{leptonterm}
L^{\mu\nu}(\ell) =  4\sqrt{2} \sum_{ij} V_{ei} V_{ej}
  \left[ { (a_{ij} m_i m_j - \ell^2 b_{ij})  \; \eta^{\mu\nu}\over
  (\ell^2 + m_i^2 - i\varepsilon ) (\ell^2 + m_j^2 - i\varepsilon) }
  \right].
\eeq

\vspace{0.08in}
\begin{center}
%%Begin InstantTeX Picture
\let\picnaturalsize=N
\def\picsize{2.3in}
\def\picfilename{onescalar.ps}
%If you do not have the picture file add: \let\nopictures=Y
%to the beginning of the file.
\ifx\nopictures Y\else{\ifx\epsfloaded Y\else\input epsf \fi
\let\epsfloaded=Y
\centerline{\ifx\picnaturalsize N\epsfxsize \picsize\fi
\epsfbox{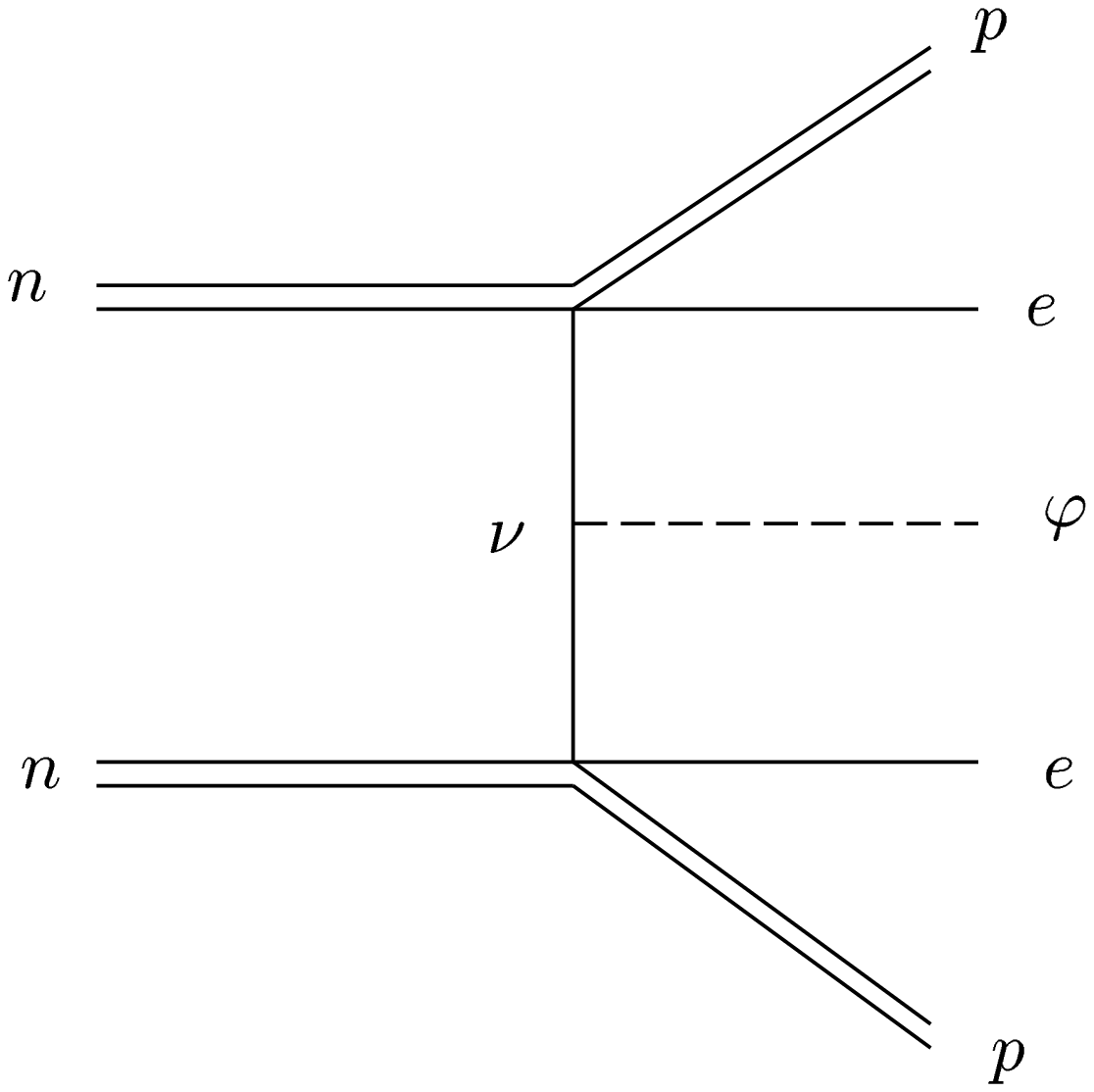}}}\fi
%%End InstantTeX Picture

\medskip
{\bf Figure 2}\\
\medskip
\end{center}

%\vspace{1cm}
\begin{quote}
{\footnotesize The Feynman graph which is responsible for $\bbm$ decay in
models for
which $\bb$ decay can arise because of sterile-neutrino exchange. }
\end{quote}

The remaining question is whether an observably large $\bbm$ decay rate can be
obtained
from these expressions without becoming in conflict with any existing
phenomenological
bounds. This can be addressed only by constructing an explicit model in which
all
of the relevant observables can be computed as a function of a common set of
underlying parameters. We therefore next present an representative model for
this
kind of theory.

An explicit sterile-neutrino model which produces $\bbm$ decays with $n=1$ is
simple to
construct\cite{CMM}. Add to the SM two electroweak-singlet, left-handed
neutrinos
$s_{\scrs \pm}$, which carry $\pm 1$ unit of an unbroken global lepton number
symmetry. Add also a singlet scalar field with lepton number $-2$. The most
general renormalizable couplings of these new particles, consistent with lepton
number conservation and the SM gauge symmetries, are
\eq
\label{ommrenmodel}
\Sc{L} = - \lambda \bar{L}H \Pr s_- - M \bar{s}_+ \Pr s_-
        -\frac{g_+}{2} \bar{s}_+ \Pr s_+ \varphi - \frac{g_- }{2}
        \bar{s}_- \Pr s_- \varphi^* + c.c.
\eeq
Here $L$ and $H$ respectively denote the usual SM lepton- and Higgs doublets,
and $\Pl$ and $\Pr$ denote the usual projections onto left- and right-handed
spinors. Majorana spinors are used throughout to represent the neutrinos,
so the spinor conjugate used above employs the charge-conjugation matrix,
$C$, according to $\ol{\nu}_i = \nu_i^\ss{T} \, C^{-1}$.

As discussed previously, the experimental absence of $\bbzn$ decay requires the
expectation value of the scalar field, $\Avg{\varphi}$, not to be larger than
$\sim 100$ keV.  Rather than fine-tuning in such a small scale, it is simpler
to
take $\Avg{\varphi} = 0$, so that lepton number is not spontaneously broken.
Then the fine-tuning simply becomes the requirement that the scalar mass
be $\lsim 10$ keV. In this case, the spectrum contains three massless
neutrinos,
$\nu_e'$, $\nu_\mu$ and $\nu_\tau$, together with a massive Dirac neutrino,
$\nu_h$, which mixes with the electron-type charged-current weak interactions.

The model can have a detectable\cite{CMM} $\bbm$ decay rate, provided that the
masses and mixings are chosen judiciously. Because the scale of the factor
$W_{\mu\nu}$
in eq.~(\ref{convolution}) is set by the nucleon momentum, $p_{\scrs N} \sim
60$
GeV, $\bb$ decay rate tends to become suppressed if all neutrino masses are
much
larger  than, or much smaller than this scale. As a result, a successful choice
of parameters, which can also avoid bounds from other laboratory experiments,
puts the heavy neutrino mass eigenstate at several hundred MeV, with a fairly
large (somewhat less than 10\%) mixing with $\nu_e$. More of the qualitative
features and problems that arise with these models are outlined in section 5,
below. Those interested in the details of the analysis are  referred to the
recent literature.\cite{CMM,BSV}

\subsection{The Case $n=3$: Two-scalar Decays}

There are two possibilities for producing $n=3$ decays: two light scalars could
be emitted without a NGB suppression of the emission amplitude, or one light
NGB-suppressed scalar could be emitted. An example of the two-scalar decay
 is given here, even though the two-scalar emission models involve the same
fine-tuning problems as do the $n=1$ models just described. The alternative
models, for which $n=3$ arises in single-scalar decays, are the subject of the
next subsection.

Consider in this case a theory of two types of left-handed neutrinos, $\nu_i$
and $N_a$, which respectively carry lepton number $L_e(\nu_i) = +1$
and $L_e(N_a) = 0$, and which are coupled to the light scalar, $\varphi$, which
has
lepton number $L_e(\varphi) = +1$. The most general renormalizable and
$L_e$-conserving Yukawa couplings involving these fields are:
\eq
\label{yukawaints}
\Sc{L}_{\rm yuk} = - \, \ol{\nu}_i \left( A_{ia} \Pl + B_{ia} \Pr
\right) N_a \; \varphi + \hc ,
\eeq
where $A_{ia}$ and $B_{ia}$ represent arbitrary Yukawa-coupling matrices.
These neutrinos are endowed with a set of generic lepton-number-conserving
masses,
 $m_{\nu_i}$ and $m_{\ss{N}_a}$. For the  $L_e = 0$ neutrinos, $N_a$, this is
accomplished by  simply  introducing a general majorana-mass term. For the
$L_e = 1$ neutrinos, however, an additional collection of $L_e = 1$
right-handed
neutrinos are required, with which $L_e$-invariant Dirac masses may be formed.

This type of theory produces $\bbmm$ decay due to the Feynman graph of Fig.~3.
Two scalars must be emitted in this decay because of $L_e$ conservation.
Evaluating this graph gives a result of the form of eqs.~\ref{spectrum},
\ref{genericrate} and \ref{convolution}, with\cite{multimajoron}:
\eq
\label{mmatrixelement}
L^{\mu\nu}(\ell) = \left( { 2 \over 3 \pi^2} \right)^{\hf} \sum_{ija} \left[
 {V_{e\nu_i} V_{e\nu_j} \Sc{N}_{ija}  \; \eta^{\mu\nu}
\over (\ell^2 + \mi^2 -i \epsilon) \, (\ell^2 + \mj^2 -i \epsilon) \,
(\ell^2 + \ma^2 -i \epsilon) } \right] ,
\eeq
where the factor, $\Sc{N}_{ija}$, denotes:
\eq
\label{numerator}
\Sc{N}_{ija} \equiv (-\ell^2) \Bigl[ A_{ia} B_{ja} \mi + A_{ja} B_{ia} \mj
+ B_{ia} B_{ja} \ma \Bigr] + A_{ia} A_{ja}  \mi \mj \ma .
\eeq

\vspace{0.08in}
\begin{center}
%%Begin InstantTeX Picture
\let\picnaturalsize=N
\def\picsize{2.3in}
\def\picfilename{twoscalar.ps}
%If you do not have the picture file add:
%\let\nopictures=Y
%to the beginning of the file.
\ifx\nopictures Y\else{\ifx\epsfloaded Y\else\input epsf \fi
\let\epsfloaded=Y
\centerline{\ifx\picnaturalsize N\epsfxsize \picsize\fi
\epsfbox{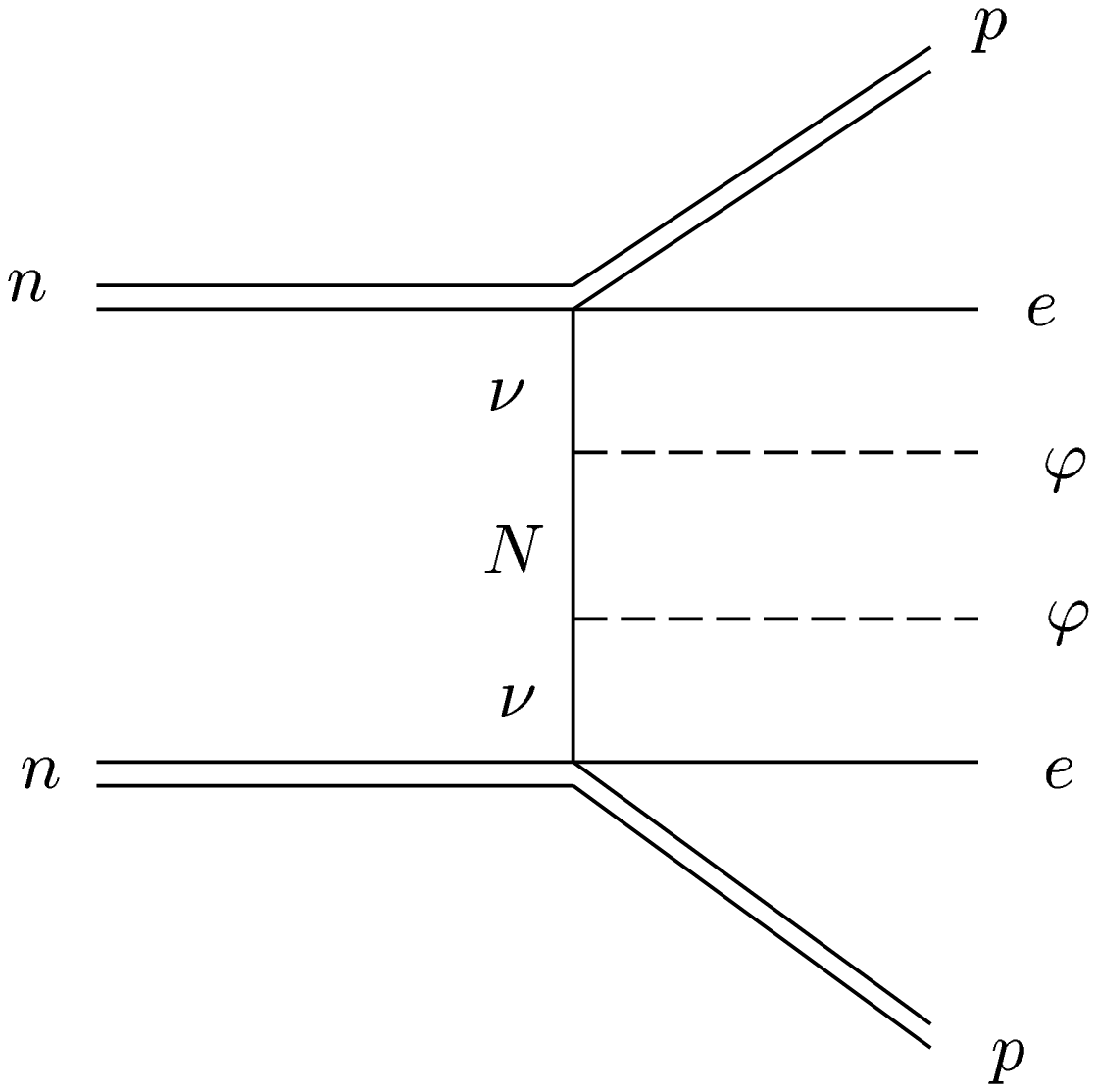}}}\fi
%%End InstantTeX Picture

\medskip
{\bf Figure 3}\\
\medskip
\end{center}

%\vspace{2cm}
\begin{quote}
{\footnotesize The Feynman graph which is responsible for $\bbmm$ decay in
models for
which $\bb$ decay arises because of sterile-neutrino exchange. }
\end{quote}

It is just possible to obtain a detectably large decay rate in this kind of
model
without running into conflict with other laboratory bounds. Because of the
comparatively soft electron spectrum (since $n=3$), the total integrated decay
rate tends to be suppressed compared to $n=1$ models by an additional two
powers of the
small ratio $Q/p_{\scrs N} \sim 10^{-1}$. The additional phase space also
introduces
additional suppression due to dimensionless factors of $1/2\pi$. As a result
the total
rate tends to be much smaller than in a comparable model for which $n=1$. This
makes it
more difficult to obtain observable decays, and a sufficiently large rate
generally
requires sterile neutrinos to lie in the mass range in the vicinity of 100 MeV
which
optimizes the $\bb$ decay rate.

\subsection{The Case $n=3$: Single-scalar Decays}

We next turn to models which predict only single-scalar $\bbm$, but for which
the
spectrum nevertheless has $n=3$. This may be ensured by constructing the model
to ensure that the emitted scalar is a NGB which carries two units of conserved
electron number.

The rate for $\bbm$ decay is given by evaluating the result for the Feynman
graph
of Fig.~2. using the generic Yukawa coupling of eq.~(\ref{yukawa}). By virtue
of
the light scalar being a NGB carrying unbroken electron number, one finds that
the
leading result in powers of the lepton momenta, as given by expression
\ref{leptonterm}, vanishes identically. It is therefore necessary to work to
next
order in these momenta, which raises the resulting spectral index from $n=1$ to
$n=3$. The resulting amplitude then satisfies $L^{\mu\nu} = - L^{\nu\mu}$,
with\cite{CMM}:
   \eqa
\label{bbcmamp}
L^{0m} &=& - 4\sqrt{2} \sum_{ij} V_{ei} V_{ej} b_{ij}
   \left[ { \ell^m \over  (\ell^2 + m_i^2 - i\varepsilon )
   (\ell^2 - m_j^2 +i\varepsilon)} \right] , \nn\\
L^{mn} &=& - 4\sqrt{2} \sum_{ij} V_{ei} V_{ej} b_{ij}
   \left[ {\epsilon^{mnr} \ell_r \over  (\ell^2 + m_i^2 - i\varepsilon )
   (\ell^2 - m_j^2 +i\varepsilon)} \right] .
\eeqa

An example of a renormalizable model for which this is the decay formula is
given\cite{CMM} by requiring the theory to have a nonabelian flavour symmetry,
$G = SU_{\scrs F}(2) \times U_{\scrs L'}(1)$ which gets broken down to the
unbroken
electron number. To implement this symmetry-breaking pattern, introduce an
electroweak-singlet scalar field, $\Phi_i$, which transforms under $G$ like
$({\bf 2}, -1)$. Also introduce the electroweak-singlet left-handed neutrino
fields, $N \sim ({\bf 2}, 0)$ and $s_{\scrs\pm} \sim ({\bf 1}, \pm 1)$.

The most general renormalizable lagrangian involving the new fields which
respects
all of the symmetries is
\eq
\label{cmmlagrangian}
 \Sc{L} =  - \lambda \bar{L}H\Pr s_-  - M \bar s_+ \Pr  s_- -
         g_+ \, (\bar{N} \Pl  s_+) \; \Phi - g_- \, (\bar{N} \Pl  s_-) \;
        \tw{\Phi} + c.c.
\eeq
$\tw{\Phi} = i \tau_2 \Phi^*$ represents the conjugate $SU_{\scrs F}(2)$
doublet,
where $\tau_2$ is the second Pauli matrix acting on flavour indices.
The scalar potential is then chosen to ensure that $\Phi$ gets a VEV, which we
assume to take the form $\Avg{ \Phi} = { 0 \choose v}$.

The soft $n=3$ spectrum suppresses the integrated decay rate, giving a
detectable
spectrum only when all couplings and masses are optimal. This once again places
the sterile neutrinos in the mass range of several hundred MeV, with
significant
mixing with the electron neutrino.

\subsection{The Case $n=7$}

As a final example, consider the case with spectral index $n=7$. This decay is
produced by ensuring that the light scalar is a NGB, and that it carries only
one
unit of electron number so that the decay rate is derivatively suppressed, and
two scalars must be emitted in the decay.

The $\bbmm$ decay rate is found by evaluating the Feynman graph of Fig.~3. Once
again, the symmetry properties make it necessary to work to higher order in
the lepton momenta than was necessary for the $n=3$ $\bbmm$ decay.  It turns
out
that the expression for the resulting amplitude is quite cumbersome when given
in terms of the couplings of eq.~(\ref{yukawaints}), and so we use instead
variables for which the derivative coupling nature of the Goldstone bosons is
manifest from the outset. The trilinear coupling to neutrinos of a Goldstone
boson
carrying $L_e = 1$, then becomes:
\eq
\label{derivyukawa}
\Sc{L}_{\rm gb} = -i \; \ol{\nu}_i \gamma^\mu ( X_{ia} \Pl +
Y_{ia} \Pr) N_a \; \partial_\mu \varphi + \hc,
\eeq
where the coefficients $X_{ia}$ and $Y_{ia}$ are coupling matrices that can be
computed in any specific model.\cite{CMM,multimajoron} Using these
interactions to evaluate the
amplitude given by the diagram of Fig.~3 gives, for the special
case\cite{multimajoron}
 $Y_{ia} = 0$:
\eq
\label{newmatrixelement}
L^{\mu\nu}(\ell) =  \left({ 4 \over 105 \pi^2} \right)^{\hf}
\; \sum_{ija} \; \left[  {V_{e\nu_i} V_{e\nu_j} \tw{\Sc{N}}_{ija}\eta^{\mu\nu}
 \over (\ell^2 + \mi^2 -i \epsilon) \,
(\ell^2 + \mj^2 -i \epsilon) \, (\ell^2 + \ma^2 -i \epsilon) } \right] \;
 .
\eeq
$\tw{\Sc{N}}_{ija}$ represents the following expression:
\eq
\label{newnumerator}
\tw{\Sc{N}}_{ija} \equiv (-\ell^2)  X_{ia} X_{ja} \ma  .
\eeq

It is fairly unlikely that this decay will be found in $\bb$ experiments in the
forseeable future, even if Nature should actually work contain $n=7$
$\bbmm$ decays. There are  two reasons why this is so.
First, the very high spectral index for this decay, $n=7$, raises several
problems for
detecting this kind of decay. Firstly, since the electron spectrum is {\em
softer} than that of the SM $\bbtn$ decay, the decay electrons tend to come out
with low energies, and so would be difficult to distinguish from background.

Secondly, the soft spectrum greatly suppresses the integrated total decay rate,
making it very difficult to get a detectable decay in a model which also
satisfies
all other laboratory bounds. Even though models which produce this spectrum
have
been constructed\cite{multimajoron}, none have been found which are both
phenomenologically viable and which predict a detectable $\bbmm$ decay rate.

\section{Models: General Features and Conclusions}

A comparison of models in these four classes leads to a number of reasonably
robust
conclusions.

\begin{enumerate}
\item
The models for which $n=1$ and those which predict $n=3$ $\bbmm$ decays
 illustrate in detail the general fine-tuning problem that was argued in
section 2
to be endemic to these kinds of $\bbm$ decay. The requirement that a light
scalar
should exist, and that $\bbzn$ should not be predicted at an unacceptably large
rate,  taken together require a fine tuning of the scalar potential to ensure
either
an extremely small scalar mass, or an equally small $L_e$-breaking \vev.
Although
supersymmetric models along these lines have been constructed\cite{oscar} which
are technically natural, they are also quite contrived and complicated.

\item
Models with softer electron spectra give smaller integrated decay rates, since
the total decay becomes suppressed by higher powers of the small endpoint
energy $Q$.
This implies that theories predicting $n=1$ decays have an easier time
producing
observably large decay rates than do models for which $n$ is larger.  As a
result these
models tend to offer the largest latitude to accomodate other laboratory limits
and
phenomenological constraints.

To produce acceptably large scalar-emitting $\bb$ decay rates, models for
which $n=3$ must typically have all of the relevant dimensionless couplings
be $O(1)$, have the mixings of the relevant sterile neutrino be as large as are
phenomenologically allowed ($\lsim 10\%$), and have the participating new
neutrino states have masses in the optimal mass range of a few hundred MeV.
This mass range is prefered since it is comparable to the typical momenta,
$p_{\scrs N}$  of the  decaying nucleons within the nucleus, and so does not
lead to suppressions of the form of $M/p_{\scrs N}$ or $p_{\scrs N}/M$.
The resulting models therefore can work, but do not leave a great deal of
freedom to accomodate other limits. Of the $n=3$ models, those which emit only
a single
scalar tend to predict larger decay rates than the two-scalar decays because
they are
not suppressed by additional small phase-space factors.

Models for which $n=7$ are the worst case, and have decay rates that
are sufficiently suppressed by powers of $Q/p_{\scrs N}$ that they are
unlikely to be experimentally detectable for the forseeable future.

\item
Models with electron spectra as soft as the $n=7$ decays are also harder to
detect
for another reason, besides the size of their total decay rate. Most of the
background in $\bb$ decay experiments occurs for electron energies which
are comparatively soft, and so it is the soft electrons which are the ones
which are hard to dig out of this background. This makes it all the more
unlikely that
these decays will turn up in the experimental data.

\item
Because the contribution of sterile neutrinos to scalar-emitting $\bb$ decays
become suppressed when their masses are much larger or much smaller than
around 100 MeV, and since couplings and mixings tend to have to be large in
order to produce an observable decay, the experimental discovery of $\bbm$
or $\bbmm$ decay  would strongly suggest the existence of new particles
in this mass range. The signals for such particles could be: anomalous bumps
in $K\to e \nu$ or $\pi \to e \nu$ decays; violations of weak universality in
leptonic $\pi$ decays, possible Zenlike monojet events at LEP, \etc. Signals in
beam dump
experiments would not necessarily be expected, since the heavy neutrinos
would dominantly decay invisibly into ordinary neutrinos and the light scalars.

\item
All models which produce scalar-emitting $\bb$ decays necessarily have at least
one very light scalar which is significantly coupled to the electron neutrino.
As a
result, all of these models generically run into trouble with Big Bang
nucleosynthesis.
The scalars tend not to decouple from the ordinary neutrinos, and so tend to
violate
the constraints on the gravitating degrees of freedom which can exist at the
epoch around $T \sim 1$ MeV.  This bound can be evaded for comparatively
special values
of the masses and couplings of the particles
involved.\cite{loophole,CMM,multimajoron}
The loophole arises  since there can be an interval during which
the neutrinos are no longer in chemical equilibrium with the protons and
neutrons,
but where the $n/p$ ratio  has still not frozen out. During this interval the
annihilation of sterile neutrinos can raise the electron neutrino abundance,
which in turn acts to deplete the neutron abundance. A sterile neutrino which
annihilates in this way effectively counts as a {\em negative} number of
neutrino species, and so can counteract the positive contribution of the light
scalars.

Of course, the details of the loophole are less interesting than the simple
fact that the
loophole exists. Although nucleosynthesis considerations generically disfavour
models
with additional light scalars, nucleosynthesis should and would be re-evaluated
should a
scalar-emitting $\bb$ decay mode be observed.

\item
The nuclear matrix elements that are required to evaluate the $\bb$ decay
rates in essentially all of the models are the same as those which have long
been studied\cite{Rosen,Haxton,KK} within the context of $\bbzn$ and
$\bbm$ decay in the GR model. This can be seen since $L^{\mu\nu}
\propto \eta^{\mu\nu}$, which implies that the decay rate depends only
on the nuclear form factor combination:  ${W^\mu}_\mu = \wgt - \wf$.
This is precisely the same combination as appears  in $\bbtn$ and $\bbzn$
decays, and so it is well studied in the literature.

The exception to this rule are those models which predict $n=3$ single-scalar
$\bbm$ decay, which instead depend on the antisymmetric part of $W_{\mu\nu}$.
The corresponding matrix elements are less well studied, with a corresponding
greater uncertainty in the predicted decay rates. (Explicit formulae for these
matrix elements in terms of nucleon operators are given in the
literature\cite{CMM}.)

\item
All of the models given here predict the same decay distribution as a function
of the opening angle of the two electrons, so this observable cannot be used to
distinguish one from another (as it can be used to distinguish decays mediated
through
right-handed currents). The only exception to this statement among the models
considered are those for which the final electrons are emitted from an $L_e =
2$ scalar
having electric charge $Q=-2$. Unfortunately the bounds on the masses of such
scalars
make the resulting $\bb$ decay rate undetectable\cite{multimajoron}.
\end{enumerate}

To summarize, $\bb$ experiments can legitimately expect to see scalar emitting
decays even though the original models which proposed this decay mode have
since been ruled out by constraints such as those coming from LEP. If such
decays are
seen, they are most likely to point to new physics with properties which are
quite different than what would have been expected from these traditional
models. Although decay spectra with $n=1, 3,5,7,...$ are all possible, the
basic combinations are $n=1,3$ and 7. The $n=7$ spectrum is very unlikely to
seen,
however, due to the strong suppression of this decay rate by powers of the
small energy
release, $Q$. Two-scalar $n=3$ decays also tend to be suppressed compared to
single-scalar decays having the same spectrum due to the presence of additional
small
phase-space factors. Furthermore, naturalness considerations (driven by the
strong
bound on the occurence of the $\bbzn$ decay mode) disfavour those models which
predict
$n=1$ $\bbm$ decays, or $n=3$ $\bbmm$ decays. Thus, from a theoretical
perspective the
$n=3$ $\bbm$ deays are the most natural. If such decays are seen we may also
probably
expect some new developments in nucleosynthesis and in precision experiments
such as
those which constrain lepton universality.

May we live to see such exciting times!

\section{Acknowledgements}

I would like to thank the workshop's organizers for
providing such a splendid setting for the workshop, and for their kind
invitation to speak. It is also a pleasure to acknowledge my collaborators
in the research described here: Peter Bamert, Jim Cline and Rabi Mohapatra. Our
funds have been provided by linear combinations of N.S.E.R.C.\ of Canada,
les Fonds F.C.A.R.\ du Qu\'ebec, the Swiss National Foundation and the
U.S. National Science Foundation.

\def\pr#1{\it Phys.~Rev.~{\bf #1}}
\def\np#1{\it Nucl.~Phys.~{\bf #1}}
\def\pl#1{\it Phys.~Lett.~{\bf #1}}
\def\prc#1#2#3{{\it Phys.~Rev.~}{\bf C#1} (19#2) #3}
\def\prd#1#2#3{{\it Phys.~Rev.~}{\bf D#1} (19#2) #3}
\def\prl#1#2#3{{\it Phys. Rev. Lett.} {\bf #1} (19#2) #3}
\def\plb#1#2#3{{\it Phys. Lett.} {\bf B#1} (19#2) #3}
\def\npb#1#2#3{{\it Nuc. Phys.} {\bf B#1} (19#2) #3}
\def\etal{{\it et.al. \/}}

\bibliographystyle{unsrt}

\begin{thebibliography}{99}
%
\bibitem{expreview}
Recent reviews of the experimental situation may be found in: M. Moe, {\it
Int.\ J.\
Mod.\ Phys.} {\bf E2} (1993) 507; M. Moe and P. Vogel, {\it Ann. Rev. of Nucl.
and
Part. Sc.} (to appear).
%
\bibitem{vectors}
C.D. Carone, \plb{308}{93}{85}.
%
\bibitem{CMM}
C.P. Burgess and J.M. Cline, \plb{298} {93}{141}; \prd{49}{94}{5925}.
%
\bibitem{icnapp}{C.P. Burgess and J.M. Cline, in the proceedings of {\it The
1st
International Conference on Nonaccelerator Physics}, Bangalore, January 1994,
(World Scientific, Singapore).}
%
\bibitem{multimajoron}
P. Bamert, C.P. Burgess and R. Mohapatra, preprint McGill-94/18, NEIP-94-010,
UMD-PP-95-78 (hep-ph/9412365)
%
\bibitem{GR}
G.B. Gelmini and M. Roncadelli, \plb{99} {81}{411}.
%
\bibitem{GGN}
H.M. Georgi, S.L. Glashow  and S. Nussinov, \npb{193} {81)}{297}.
%
\bibitem{twoscalar}
R. Mohapatra and E. Takasugi,\plb{211}{88}{192}.
%
\bibitem{oscar}C.P. Burgess and O. Hern\'andez, \prd{48}{93}{4326}.
%The Russian collaborators?.
%
\bibitem{CMP}
Y. Chikashige,  R.N. Mohapatra and R.D. Peccei, \prl{45}{80}{1926}.
%
\bibitem{expgeff}
See e.g. M. Beck {\it et.al.}, \prl{70}{93}{2853} and
references therein, or J.-L. Vuilleumier {\it et.al.}, {\it Nucl.
Phys. (Proc. Suppl.)} {\bf B31}{ (1993)}.
%
\bibitem{kai}
Kai Zuber, in Relativistic Astrophysics and Particle Cosmology, ed.
by C.W. Akerlof and M.A. Srednicki, Annals of the New York Academy of Sciences,
Vol. 688 (New York, 1993).
%
\bibitem{heid}
 A. Balysh, et al., Heidelberg preprint, hep-ex/9502007 (July 1994).
 %
\bibitem{DoiTomoda}
M. Doi, T. Kotani and E. Takasugi, {\it Prog. Theor. Phys.}~Suppl. {\bf
83} (1985) 1; T. Tomoda, {\it Rept. Prog. Phys.}~{\bf 54} (1991) 53.
%
\bibitem{Haxton}
See also: W.C. Haxton and G.J. Stevenson, Progress in Particle and
\npb{12} {84}{409}; T. Tomoda, A. Faessler and K.W. Schmid,
{\it Nucl. Phys.} {\bf A452} (1986) 591.
%
\bibitem{Rosen}
A. Halprin, P. Minkowski, H. Primakoff and P. Rosen, \prd{13}{76}{2567};
S.P. Rosen, Arlington preprint UTAPHY-HEP-4 (1992); and references therein.
%
\bibitem{KK}
H. Klapdor-Kleingrothaus, K. Muto and A. Staudt, {\it Europhys.
Lett.} {\bf 13}, 31 (1990); M. Hirsch et. al. {\it. Zeit. Phys. A}
{\bf 345}, 163 (1994);
H. Klapdor-Kleingrothaus , {\it Prog. Part. Nucl. Phys.} {\bf 32},
 261 (1994) for a review.
%
\bibitem{BSV}
Z.G. Berezhiani, A.Yu. Smirnov and J.W.F. Valle, \plb{291}{92}{99}.
%
\bibitem{loophole}
K.~Enqvist, K.~Kainulainen and M.~Thomson, \prl{68}{92}{744}.
\end{thebibliography}

\end{document}